# Coordinated Frequency Control through Safe Reinforcement Learning

Yi Zhou[1], Liangcai Zhou[1], Di Shi[2], Xiaoying Zhao[2]
[1]East China Branch of State Grid Corporation, Shanghai, China
[2]AINERGY, Santa Clara, CA, USA
Email: di.shi@ainergysolutions.com

*Abstract*—With widespread deployment of renewables, the electric power grids are experiencing increasing dynamics and uncertainties, with its secure operation being threatened. Existing frequency control schemes based on day-ahead offline analysis and minute-level online sensitivity calculations are difficult to adapt to rapidly changing system states. In particular, they are unable to facilitate coordinated control of system frequency and power flows. A refined approach and tools are urgently needed to assist system operators to make timely decisions. This paper proposes a novel model-free coordinated frequency control framework based on safe reinforcement learning, with multiple control objectives considered. The load frequency control problem is modeled as a constrained Markov decision process, which can be solved by an AI agent continuously interacting with the grid to achieve sub-second decision making. Extensive numerical experiments conducted at East China Power Grid demonstrate the effectiveness and promise of the proposed method.

## I. Introduction

Increasing penetration of renewables and hybrid operation of UHV AC/DC transmission networks have significantly increased system uncertainties and dynamics at East China Power Grid, posing new challenges to its frequency security. Recent events such as *Jinsu* HVDC line bipolar blocking and *Binjin* HVDC line unipolar blocking in 2015 both caused considerable frequency drop, raising warnings in frequency control of interconnected power grids [1].

Presently, the East China Power Grid adopts a frequency control system featured by dynamic area control error (ACE) [2]. When a disturbance occurs, five regional (provincial) systems, each working as a control area, will share the power imbalance based on prescribed proportions of the spinning reserves to recover system frequency. In practice, the use of preset apportionment ratio often causes the flows on tie-lines and flowgates to exceed limits. The speed and accuracy of the existing model-based control framework based on day-ahead offline analysis and sensitivity coefficients are insufficient, and consequently the control and adjustment process need to be iterated multiple times before arriving at feasible states. Therefore, there are pressing needs for a more refined frequency and power flow coordinated control approach which can assist dispatchers in making decisions online.

In recent years, deep reinforcement learning (DRL) has shown promise in solving complex problems in various fields, including power systems [3]-[6]. As one of the pioneering works, Diao et al. proposed an autonomous voltage control (AVC) framework based on DRL which gives system dispatchers instructions in sub-seconds [7]. Since then, various versions of DRL algorithms have been applied to low frequency oscillation damping control [8], reactive power dispatch [9], short-term load forecasting [10], load model identification [11], autonomous line flow control [12], network topology optimization [13], and AC OPFs [14].

In the area of frequency control, Yan et al. proposed a deep deterministic policy gradient (DDPG) based approach with continuous action space for a single-area system [15], Adibi et al. applied an actor-critic algorithm to microgrid frequency control [16], and Rozada et al. proposed a distributed load frequency control algorithm based on DDPG [17]. The existing works suffer from deficiencies from two aspects. First, they mainly consider a single control objective of fast frequency recovery using pre-determined coefficients for power distribution among generators. They may work well for small systems and microgrids, but for large-scale systems with multiple control areas, the aforementioned problem of overflows on tie-lines become obvious. Second, few existing works consider the safety issues of agents during the action searching process, which is crucial for mission critical applications such as power system frequency control.

To address the gap, this paper proposes a coordinated frequency control framework based on safe reinforcement learning with multiple control objectives and operational constraints considered. The proposed framework targets applications at the secondary and tertiary frequency control loops while works with the existing primary frequency control. The following contributions are made: 1) safety is taken into account during the action search process so that various operational constraints are satisfied while agent interacts with the environment; 2) power sharing among generators are coordinated and dynamically adjusted to ensure power flows on tie-lines and flowgates stay within specified ranges; 3) the proposed approach has been validated and demonstrated on East China Power Grid, one of the largest interconnected power systems with multiple HVDC lines.

## II. Formulations and Proposed Framework

### A. DRL Basics

Reinforcement learning (RL) has shown exceptional success in solving sequential decision problems of impressive difficulty by optimizing a return through trial and error. The objective of maximizing this cumulative expected return is achieved by continuously interacting with the environment, as depicted in Fig. 1.

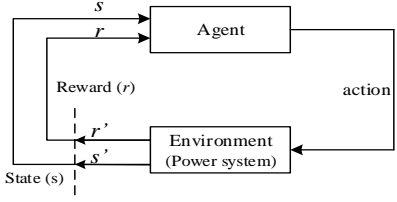

Fig. 1. An RL agent interacts with the environment.

Each time the power grid performs an action given by the agent, it returns a new system state and calculates the corresponding reward; and the agent will learn and improve the action strategy in the process of continuously interacting with the grid. Combining deep learning (DL) with RL defines the fields of DRL, in which the deterministic or stochastic policy is usually approximated by a deep neural network.

It is worth mentioning that the environment can either be the power system itself or its high-fidelity simulator.

### B. Constrained Markov Decision Process (CMDP)

The CMDP can be described as a five-dimension tuple ($S$, $A$, $P_a$, $R_a$, $C$), where $S$ represents the state space, $A$ is action space, $P_a(s, s')=Pr(s_{t+1}=s'|s_t=s, a_t=a)$ is the probability of state transition from $s_t$ to $s_{t+1}$ after taking action $a_t$, $R_t(s,s')$ is the reward obtained for the transition, and $C$ represents a set of constraints.

Each set of trajectories that constrain the Markov decision process corresponds to a (discounted) return. The solution goal is to obtain a control strategy $\pi$, so that the system state meets the constraining conditions and the expected return $J(\pi)$ is the largest through execution of the strategy, as defined below.

$$\max_{\pi} E\left[\sum_{t=0}^{T} \gamma^t R_t\right] \quad s.t. \ C^{\pi}(s) \leq \bar{C} \quad (1)$$

where $C^{\pi}(s)$ is the cost function, $\gamma$ is the discount factor, $\bar{C}$ is the upper limit of cost function, and $R_t$ is short for $R_{a_t}(s,s')$.

A state value function which depends upon the initial system state is written as:

$$V^{\pi}(s) = E_{\tau \sim \pi}\left[\sum_{t=0}^{T} \gamma^t R_t | s\right] \quad (2)$$

where $\tau$ is a trajectory. Similarly, an action value function $Q^{\pi}(s,a)$ can be defined as:

$$Q^{\pi}(s,a) = E_{\tau \sim \pi}\left[\sum_{t=0}^{T} \gamma^t R_t | s, a\right] \quad (3)$$

Both state and action value functions satisfy the Bellman equation [18]:

$$V^{\pi}(s_t) = E_{\substack{a_t \sim \pi \\ s_{t+1} \sim P_a}}[R_t + \gamma V^{\pi}(s_{t+1})] \quad (4)$$

$$Q^{\pi}(s_t, a_t) = E_{\substack{a_{t+1} \sim \pi \\ s_{t+1} \sim P_a}}[R_t + \gamma Q^{\pi}(s_{t+1}, a_{t+1})] \quad (5)$$

### C. Safety-Constrained Soft Actor-Critic (SSAC)

For power system frequency control, collecting a large number of effective training data samples is often costly, and therefore off-policy algorithms with higher sampling efficiency are generally preferred. Among various off-policy RL algorithms, SAC is adopted in this work considering its superior performance, which maximizes the expected return by exploring as many control actions as possible, leading to a better chance of finding the optimum [19]. In this work, a novel algorithm is developed by extending SAC to include operational safety constraints. Essentially, we try to restrict the entire policy space to a smaller region so that the operational constraints on the power flows of tie-lines and flowgates are satisfied, as illustrated in Fig. 2.

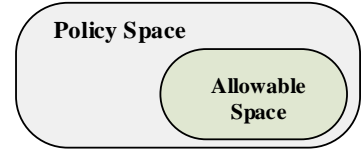

Fig. 2. Policy space and the set of allowable policies.

In SAC, the control strategy maximizes the sum of expected return and entropy, and the optimal policy can be written as:

$$\pi^* = \arg\max_{\pi} E_{\tau \sim \pi}\left[\sum_{t=0}^{T} \gamma^t (R(s_t, a_t) + \alpha H(\pi(\cdot|s_t)))\right] \quad (6)$$

where $H(\pi(\cdot|s_t))$ is the entropy of policy $\pi$ at state $s_t$, $\alpha$ is a trade-off coefficient (temperature parameter). The two value functions $V^{\pi}(s_t)$ and $Q^{\pi}(s_t, a_t)$ follow:

$$V^{\pi}(s_t) = E_{a_t \sim \pi}[Q^{\pi}(s_t, a_t)] + \alpha H(\pi(\cdot|s_t)) \quad (7)$$

The training of an SAC agent is similar to other gradient descent algorithms. To evaluate the control policy, a deep neural network with stochastic gradient can be utilized. For the two value functions $V_{\psi}^{\pi}(s_t)$ and $Q_{\theta}^{\pi}(s_t, a_t)$, parameters of the corresponding neural networks are represented as $\psi$ and $\theta$, respectively, and parameters of the policy network $\pi_{\phi}(a_t, s_t)$ are represented by $\phi$. In SAC, two sets of value functions are utilized with one called "soft" value functions, which are updated periodically to improve the stability and reliability of the algorithm. The soft state value function is updated by minimizing the residual shown below.

$$L_V(\psi) = E_{s_t \sim D}\left[\frac{1}{2}\left(V_{\psi}^{\pi}(s_t) - E_{a_t \sim \pi_{\theta}}[Q_{\theta}^{\pi}(s_t, a_t) - \alpha \log \pi_{\phi}(a_t|s_t)]\right)^2\right] \quad (8)$$

where $D$ is the distribution of sampled data. The gradient of (8) is calculated as:

$$\hat{\nabla}_{\psi} L_V(\psi) = \nabla_{\psi} V_{\psi}^{\pi}(s_t)[V_{\psi}^{\pi}(s_t) - Q_{\theta}^{\pi}(s_t, a_t) + \alpha \log \pi_{\phi}(a_t|s_t)] \quad (9)$$

Similarly, parameters of the soft action value $Q$ function can be updated by minimizing:

$$L_Q(\theta) = E_{(s_t, a_t) \sim D}\left[\frac{1}{2}\left(Q_{\theta}^{\pi}(s_t, a_t) - \hat{Q}(s_t, a_t)\right)^2\right]$$

$$\hat{Q}(s_t, a_t) = R(s_t, a_t) + \gamma E_{s_{t+1} \sim p}\left[V_{\bar{\psi}}^{\pi}(s_{t+1})\right] \quad (10)$$

where $\bar{\psi}$ is the moving average of $\psi$. Solution of (10) can be obtained via iterating using the following gradient:

$$\hat{\nabla}_{\theta} L_Q(\theta) = \nabla_{\theta} Q_{\theta}^{\pi}(s_t, a_t)[Q_{\theta}^{\pi}(s_t, a_t) - R(s_t, a_t) - \gamma V_{\bar{\psi}}^{\pi}(s_{t+1})] \quad (11)$$

The policy network can be updated by minimizing the Kullback-Leibler (KL) divergence, as:

$$L_\pi(\phi) = E_{s_t \sim D}\left[E_{a_t \sim \pi_\phi}\left[\alpha \log\left(\pi_\phi(a_t|s_t)\right) - Q_\theta^\pi(s_t, a_t)\right]\right] \quad (12)$$

$$\hat{\nabla}_\phi L_\pi(\phi) = \alpha \nabla_\phi \log\left(\pi_\phi(a_t|s_t)\right) + \left(\alpha \nabla_{a_t} \log\left(\pi_\phi(a_t|s_t)\right) - \nabla_{a_t} Q_\theta^\pi(s_t, a_t)\right) \nabla_\phi f_\phi(\varepsilon_t; s_t)$$

where $\nabla_{a_t}$ is the gradient of action $a_t$, $\varepsilon_t$ is error of the input vector, $a_t$ can be obtained from the transformation of neural network $f_\phi(\varepsilon_t; s_t)$.

During the agent training process, before and after an action is executed, all transmission line power flows should stay within the ratings or constrained ranges, that is, $F_t \leq F_{limit}$, where the two variables are the line flows at time $t$ and the corresponding limits. We introduce one type of cumulative constraint in this work. The instantaneous ones can be formulated in a similar manner. According to (1), the cost function is selected as the number of lines whose power flows are over the limits, as:

$$C^\pi(s) = E_{\tau \sim \pi}\left[\sum_{t=0}^T \gamma^t c_t(s)\right]$$
$$= E_{\tau \sim \pi}\left[\sum_{t=0}^T \gamma^t \left(\sum_{i=1}^K \mathbb{1}(F_{t+1}^i > F_{limit}^i)\right)\right] \quad (13)$$
$$\bar{C} = \frac{1-\gamma^T}{1-\gamma}\bar{c}$$

where $c_t(s)$ and $\bar{c}$ are the cost function at time $t$ and the corresponding upper limit, $\mathbb{1}(\cdot)$ is an indicator function, and $K$ is the total number of lines. Other types of constraints can be enforced in a similar manner.

Combing (1), (6) and (13), the CMDP of the proposed coordinated frequency control problem can be developed as follows with $D$ denoting a data buffer storing historical operational data:

$$\max_\pi E_{s \sim D}\left[\sum_{t=0}^T \gamma^t (R(s_t, a_t) + \alpha H(\pi(\cdot|s_t)))\right]$$
$$s.t. \ E_{s \sim D}\left[\sum_{t=0}^T \gamma^t c_t(s)\right] \leq \bar{C} \quad (14)$$

The corresponding Lagrangian is:

$$\mathcal{L}(\pi, \lambda) = E_{s \sim D}\left[\sum_{t=0}^T \gamma^t (R(s_t, a_t) + \alpha H(\pi(\cdot|s_t)))\right] + \lambda\left(\bar{C} - E_{s \sim D}\left[\sum_{t=0}^T \gamma^t c_t(s)\right]\right) \quad (15)$$

which can be solved using the Lagrange multiplier method, and $\lambda$ can be solved through iteration with a given initial value:

$$\lambda^{k+1} = \lambda^k - \sigma_\lambda \nabla_\lambda \mathcal{L} \quad (16)$$

Parameters of the two value function networks and the policy network can be updated iteratively, in a similar way as (8)-(12), and therefore will not be discussed here for the interest of space.

## III. DESIGN OF THE AGENT

The key elements of training SSAC agents for multi-objective coordinated frequency control are discussed in this section.

### 1) Environment and Data Samples

The environment used to train an agent can either be a real power system or its simulator. The simulation environment can be the state estimation module and power flow module of the EMS, or PSD-BPA, PSS/E, and DSATools™. The data exchange between the agent and the environment can be system snapshots in standardized formats, or WAMS and SCADA data. In particular, the agent should be able to obtain frequency deviation and power imbalance from the grid environment (i.e., dynamic ACE system of the East China Grid).

As effective data samples are critical for training SSAC agents, they should be collected in a way to be representative of a wide range of system operating conditions. When a major change occurs in the topology of the grid, the changes are reflected in historical system snapshots to be used for the training. If the agent is trained for a possible future operating condition with major topology changes, the change needs to be reflected in the data samples as well.

### 2) State and Action Spaces

The system states include bus voltage $V$, line flows ($P_L$, $Q_L$), generator outputs ($P_g$, $Q_g$), system power imbalance/loss $P_{loss}$ or frequency deviation. The system state is defined as:

$$S = [V_1, \ldots, V_N, P_{L_1}, Q_{L_1}, \ldots, P_{L_K}, Q_{L_K}, \ldots, P_{g_1}, Q_{g_1}, \ldots, P_{g_n}, Q_{g_n}, P_{loss}] \quad (17)$$

where $N$, $K$, and $n$ are the number of buses, transmission lines, and generators in the system, respectively.

Control actions considered include adjusting generator outputs, shedding load under emergency conditions, adjusting power imports/exports from external regions through HVDC lines, etc. As adjusting HVDC imports or exports for frequency control rarely happened at East China Grid, and without losing generality, only the first two types of actions are considered, as defined below, with $m$ representing the number of loads which can be controlled/shed:

$$A = [\Delta P_{g_1}, \Delta P_{g_2}, \ldots, \Delta P_{g_n}, \Delta P_{L_1}, \ldots, \Delta P_{L_m}] \quad (18)$$

### 3) Reward Design

Definition of the reward will largely determine the performance of a SSAC agent. It is worth mentioning that getting a good reward function needs substantial engineering efforts. The proposed framework can fulfill multiple control objectives. In this work, the objective is to realize fast frequency recovery with the least cost while make sure the power flows on all tie-lines and flowgates stay within the corresponding thermal ratings or stability limits, which are major concerns of system dispatchers at East China Grid. The reward function is defined as follows:

$$C_{sys} = \sum_{i=1}^n C_{g_i}(P_{g_i}) \quad (19)$$

$$or \quad C_{sys} = \sum_{i=1}^{K}(P_i^{from} + P_i^{to})$$

$$D_{overflow} = \sum_{i=1}^{l}[max(P_{inter\_i} - P_{inter_i}^{limit}, 0)]^2 \quad (20)$$

$$D_p = \left|\sum_{i=1}^{n}\Delta P_{g_i} - \sum_{i=1}^{m}\Delta P_{L_i} - P_{loss}\right| \quad (21)$$

$$R_t = \begin{cases} E_1 - \dfrac{C_{sys}}{E_2} & if\ D_{overflow} = 0 \\ -\dfrac{D_{overflow}}{E_3} - \dfrac{D_p}{E_4} & if\ D_{overflow} > 0 \end{cases} \quad (22)$$

where $C_{sys}$ is the production cost or system loss, $C_{g_i}(\cdot)$ is the cost function of the $i$th generator, $P_i^{from}$ and $P_i^{to}$ are the active power measured at the sending and receiving ends of the $i$th line, respectively, $l$ is the number of key tie-lines or flowgates being monitored, $P_{inter\_i}$ and $P_{inter_i}^{limit}$ are the power flow through the $i$th tie-line or flowgate and its corresponding upper limit, $D_{overflow}$ is the sum of squares of overflows, and $D_p$ is system power imbalance, $E_1$ is a positive offset coefficient which makes the reward positive when the corresponding conditions are met, and $E_2$-$E_4$ are positive constants which adjust the ratios of different components and limit the range of the reward the agent can obtain. It is worth mentioning that the reward definition can be further modified to differentiate generators based on their characteristics.

*4) SSAC Training Algorithm*

The training process of an SSAC agent for coordinated frequency control is described in Algorithm I.

**Algorithm I**: SSAC Training Algorithm for Coordinated Freq. Ctrl.
**Initialize**: policy network $\phi$, value networks $\psi$ and $\theta$, target network $\bar\psi$, Lagrange multiplier $\lambda$, replay buffer $D$, smoothness factor $\tau$, $\alpha$ and $\gamma$
**for** all training episodes **do**
  **for** each environment step **do**
    observe system states and calculate action $a_t \sim \pi(\cdot|s_t)$
    execute $a_t$, observe $s_{t+1}$, $r_t$, $c_t$, and a terminating signal *done*
    store tuple $D=D\cup<s_t,a_t,r_t,c_t,s_{t+1},done>$
  **end for**
  **if** requirement is met for updating neural networks
    **for** each gradient step **do**
      randomly sample a mini-batch $M$ from $D$
      update action value network $\theta$
      update state value network $\psi$
      update policy network $\phi$
      update target network $\bar\psi \leftarrow \tau\psi + (1-\tau)\bar\psi$
      update Lagrange multiplier $\lambda$
    **end for**
**end for**
Save agent model and write logs

## IV. NUMERICAL RESULTS ON EAST CHINA GRID

The proposed framework has been validated using real operational data of East China Grid (colored portion in Fig. 3 with each color indicating a control area), one of the most heavily loaded regional power grids in the world. System operational data between May and August of 2021 are collected which consist of 4463 buses (among which 2913 are 220kV and above), 5686 transmission lines, 2051 transformers, 216 power plants (579 generators), and 49 key inter-area tie-lines/flowgates. In order to generate more representative data samples for training and testing, system loads are perturbed randomly between 90%-110%, using a similar approach as discussed in [9]. In addition, contingencies including single-pole blocking and double-pole blocking are considered and implemented to two HVDC lines named *Jinsu* and *Lingshao*. A total of 120,000 system snapshots are used for numerical experiments. Considering the engineering practice at East China Grid, the main control measure under consideration is to adjust outputs of generators and avoid shedding load unless necessary. Control decisions are made at the power plant level and generators within each power plant are adjusted proportionally based on their reserves/capacities.

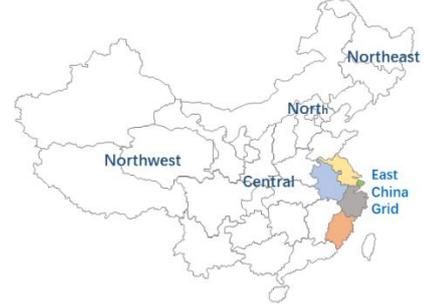

Fig. 3. Service territory of East China Power Grid.

The parameters used for training the coordinated frequency control agents are shown in Table I.

TABLE I PARAMETER SETTINGS FOR TRAINING AND TESTING THE AGENT

| | |
|---|---|
| Number of Hidden layers | 3 |
| Size of Hidden Layers | (2048, 1024, 512) |
| Batch size | 256 |
| Learning rate | 0.001 |
| Discount factor, $\gamma$ | 0.99 |
| Temperature parameter, $\alpha$ | 0.006 |
| Maximum entropy | 0.1 |
| Initial Value of Lagrange Multiplier ($\lambda$) | 0.0 |
| Size of Replay buffer, $D$ | 50,000 |
| Activation Function of Hidden Layer | ReLU |
| Optimizer | Adam |
| Smoothness Factor | 0.0002 |

These samples are divided into two portions, with the first 100,000 samples used for training and the remaining 20,000 samples for testing. Difference between the training and testing phases lie in 1) at the end of the training phase, parameters of all networks are fixed for testing; 2) the deterministic policy is used during testing while a stochastic policy is used during training.

Fig. 4 shows the average total return of evaluation rollouts during training for the proposed SSAC algorithm. We train five different instances of the SSAC algorithm with different random seeds (8, 10, 18, 22, 28), with each performing one evaluation rollout every 500 environment steps. The solid curve corresponds to the mean and the shaded region to the 99% confidence interval ($3\sigma$). The figure shows initially the return is negative and very quickly the return become positive. According to the reward definition, a positive return indicates the control goals have been achieved. Over the period of training, the total return consistently improves and its standard deviation reduces.

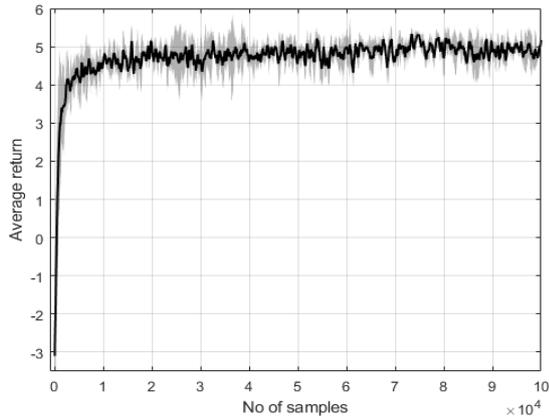

Fig. 4. Average total return during the training phase.

The average total return from the testing phase is shown in Fig. 5. Details of the testing phase for each of the five instances or agents are summarized in Table II. As the table shows, for the first two agents, both system frequency and overflows through tie-lines or flowgates are solved 100%. For the last three agents, all frequency issues are solved 100% but there are 68, 86, and 16 cases, respectively, which still have overflow problems after taking control action from the agents. The success rates for the last three agents are still very good, ranging between 99.57% to 99.66%. For all five instances, it takes the agents less than 20 *ms* to come up with a control decision.

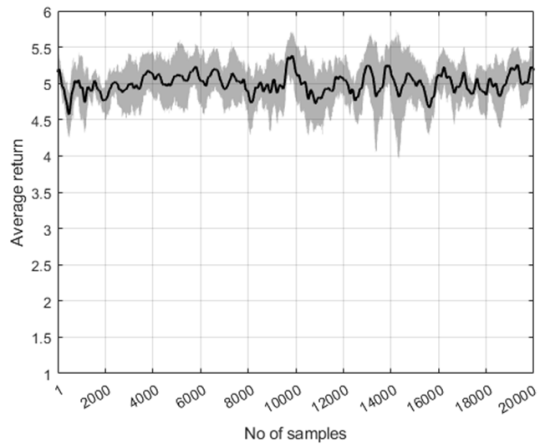

Fig. 5. Average total return during the testing phase.

TABLE II SUMMARY OF TESTING RESULTS

| Agent No. | Total cases | Unsolved cases | | Success rate (%) | Avg. decision time (ms)* |
|---|---|---|---|---|---|
| | | *w.r.t.* freq. | *w.r.t.* flows | | |
| 1 | 20,000 | 0 | 0 | 100 | 15.178 |
| 2 | 20,000 | 0 | 0 | 100 | 16.290 |
| 3 | 20,000 | 0 | 68 | 99.66 | 19.703 |
| 4 | 20,000 | 0 | 86 | 99.57 | 16.842 |
| 5 | 20,000 | 0 | 16 | 99.92 | 17.633 |

*Intel i9-7920 CPU@2.9GHz, 128GB RAM, Ubuntu 20.04.2LTS, 4×Nvidia Titan V

In order to further evaluate performance of trained agents, we compare the testing results against the case that all generators adjust their outputs to compensate system power imbalance proportionally based on their reserves, which is set as the benchmark case in the following discussion. We are particularly interested in the comparison of numbers of tie-lines/flowgates with overflows and differences in system losses. Table III summarizes the statistics of number of snapshots with different numbers of overflows. As the table shows, if generators are adjusted based on their reserves, all 20,000 cases have 1 to 3 overflows on tie-lines or flowgates. As an extension of the discussion regarding Table II, with the proposed framework, these overflows can be resolved using the trained SSAC agents, with the first two agents achieving perfect results and the last three achieving success rates greater than 99.57%.

TABLE III STATISTICS OF NUMBERS OF SNAPSHOTS WITH OVERFLOWS

| Num. of Overflows / Agent | 0 | 1 | 2 | 3 |
|---|---|---|---|---|
| Benchmark | - | 17213 | 2434 | 353 |
| Agent 1 | 20000 | - | - | - |
| Agent 2 | 20000 | - | - | - |
| Agent 3 | 19932 | 68 | - | - |
| Agent 4 | 19914 | 74 | 11 | 1 |
| Agent 5 | 19985 | 15 | 1 | - |

The improvement on system loss is evaluated based on:
$$diff_{loss}(\%) = \frac{Loss_2 - Loss_1}{Loss_2} \times 100\% \quad (23)$$
where $Loss_1$ and $Loss_2$ are system losses obtained by the agent and from the benchmark case, respectively. Differences in system loss for all five agents are shown in Fig. 6. It is observed that all five agents have improved system losses, which verifies that the loss reduction control objective has been achieved. As compared to the benchmark case, the first two agents result in more loss reduction than the remaining three. The average loss improvements for the five agents are 3.033%, 3.071%, 1.826%, 2.556%, and 2.801%, respectively.

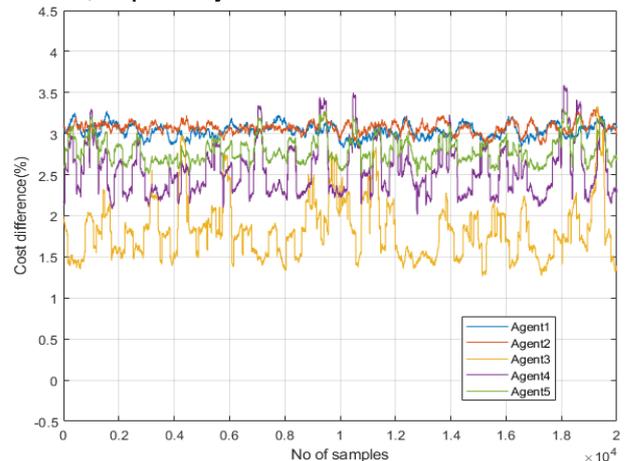

Fig. 6. Difference in system loss during the testing phase.

To sum up, the proposed framework has been demonstrated and validated at one of China's major regional power systems with promising results observed. A trained SSAC agent can master the coordinated frequency control problem starting from scratch and make control decisions in milliseconds, showing great potential in assisting system dispatcher to make decisions to ensure power system frequency security. Last but not the least, it is worth mentioning that due to the statistical and random nature of the stochastic policy in SSAC, training agents with exactly the same parameters, as this section shows, may still lead to agents with different performance. Therefore, it is always a good idea to train multiple agents, evaluate and compare their performance and chose one or more with better performance. And for online inferencing, it is desired to have multiple agents generating concurrent decisions and synthesize and/or validate them before execution.

## V. CONCLUSION AND FUTURE WORK

A coordinated frequency control framework is proposed based on safe reinforcement learning. Multiple control objectives can be achieved with guaranteed satisfaction of system operational constraints. Through extensive numerical experiments, the proposed framework has demonstrated promise at China East Grid. As the next step, we will investigate approaches that can further improve performance of the SSAC agents and methodologies that coordinate the training and inferencing processes to further automate the software packages being developed and better handle the issue of model drifting against time and major system operational (including topological) changes. We will also investigate different reward design methodologies that can differentiate generators (e.g., energy storage systems) in the decision-making process.